\documentclass[epsfig,graphics,twocolumn,showpacs,floatfix,mathbbm,prl,superscriptaddress]{revtex4}
\usepackage{amsmath,amssymb,graphicx,color}
\usepackage{float}
\usepackage{dsfont}
\newcommand{\PrYSO}{Pr$^{3+}$:Y$_2$Si{O$_5$}}
\newcommand{\YSO}{Y$_2$Si{O$_5$}}

\newcommand{\1}[1]{\mathds{1}}

\begin{document}
\title{Time Entanglement between a Photon and a Spin Wave in a Multimode Solid-state Quantum Memory}
\author{Kutlu Kutluer}
\author{Emanuele Distante}
\author{Bernardo Casabone}
\author{Stefano Duranti}
\author{Margherita Mazzera}
\affiliation{ICFO-Institut de Ciencies Fotoniques, The Barcelona Institute of Science and Technology, Mediterranean Technology Park, 08860 Castelldefels (Barcelona), Spain}
\author{Hugues de Riedmatten}
\affiliation{ICFO-Institut de Ciencies Fotoniques, The Barcelona Institute of Science and Technology, Mediterranean Technology Park, 08860 Castelldefels (Barcelona), Spain}
\affiliation{ICREA-Instituci\'{o} Catalana de Recerca i Estudis Avan\c cats, 08015 Barcelona, Spain}

\date{\today}

\begin{abstract}
The generation and distribution of entanglement are key resources in quantum repeater schemes. Temporally multiplexed systems offer time-bin encoding of quantum information which provides robustness against decoherence in fibers, crucial in long distance communication. Here we demonstrate the direct generation of entanglement in time between a photon and a collective spin excitation in a rare earth ion doped ensemble. We analyze the entanglement by mapping the atomic excitation onto a photonic qubit and by using time-bin qubits analyzers implemented with another doped crystal using the atomic frequency comb technique. Our results provide a solid-state source of entangled photons with embedded quantum memory. Moreover, the quality of the entanglement is high enough to enable a violation of a Bell inequality by more than two standard deviations.
\end{abstract}

\maketitle

Light-matter entanglement is an important resource in quantum information science. It enables complementing the advantages of using photons as flying qubits in quantum communication schemes with those of matter qubits, which are ideal for quantum storage and processing \cite{Kimble2008, Sangouard2011}. It can be achieved, for example, by interfacing quantum sources of entangled photons with long lived quantum memories \cite{Simon2007,Clausen2011,Saglamyurek2011}.
But the direct generation of light-matter entanglement, without the use of external photon pair sources, is particularly attractive in view of practical application as it generally features less complexity and can lead to higher efficiency than the so-called read-write memory protocols \cite{Afzelius2015}.

A very convenient method to directly generate light-matter entanglement in atomic ensembles is the Duan-Lukin-Cirac-Zoller (DLCZ) protocol \cite{Duan2001}. It is based on the off-resonant excitation by means of weak classical write pulses of an atomic ensemble with a lambda-system. With a small probability, a Raman scattered write photon creates a collective spin excitation, heralded by the emission of a Stokes photon. The spin wave can be converted into a second photon, the anti-Stokes photon, with the use of a strong on-resonant read pulse. In this way, the light-matter entanglement is mapped into photonic entanglement, which can be analyzed with photonic qubit analyzers. Several types of entanglement have been demonstrated using the DLCZ scheme in atomic gases, such as polarization \cite{Matsukevich2005,Riedmatten2006}, spatial modes \cite{Chen2007, Pu2017, Chrapkiewicz2017}, orbital angular momentum \cite{Inoue2006}, and time-bin \cite{Farrera2018}. The DLCZ scheme has also been demonstrated with nanomechanical resonators \cite{Riedinger2016}

The use of atomic ensembles embedded in solid matrices, such as rare earth ion doped (REID) crystals, offers numerous advantages as the coherence times are comparable to those of cold atomic clouds but the natural trapping greatly simplifies the experimental setups. Moreover, the inhomogeneous broadening of the atomic transitions can be used as a resource for quantum information multiplexing \cite{Afzelius2009,Sinclair2014}. However, the standard off resonant DLCZ scheme is difficult to apply to REID crystals, due to the very weak dipole moments of the optical transition. Alternative schemes have been proposed that use resonant excitation and rephasing techniques to counteract the inhomogeneous dephasing of the atomic dipoles \cite{Sekatski2011, Ledingham2010}. Very few attempts of implementing the DLCZ-like scheme in REID crystals have been done, demonstrating continuous variable entanglement \cite{Ledingham2012, Ferguson2016} and quantum correlation between photons and spin waves \cite{Kutluer2017,Laplane2017}. The latter demonstrations combined the DLCZ protocol and the atomic frequency comb (AFC) storage scheme \cite{Afzelius2009}.

\begin{figure*}
	\centering{\includegraphics[width=2\columnwidth]{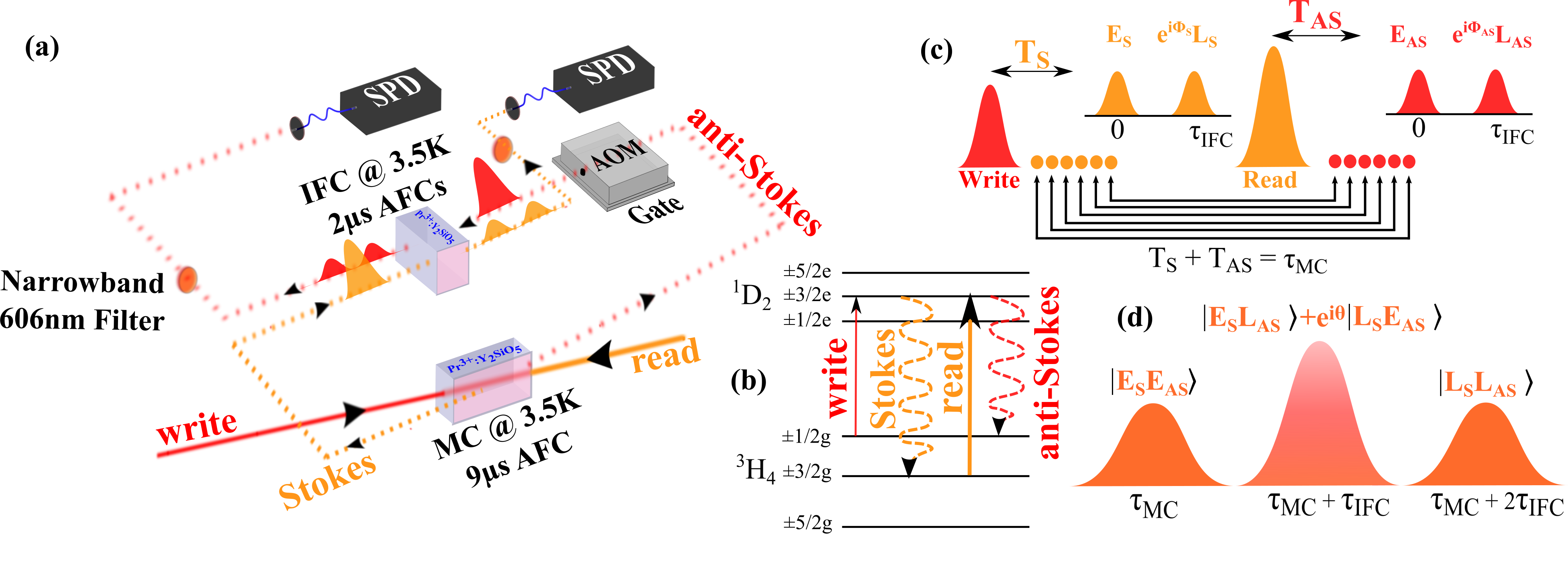}}
	\caption{
(a) Experimental setup. The write and read pulses are polarized parallel to the D$_{2}$ memory crystal (MC) axis to maximize the interaction. Both Stokes and anti-Stokes photons pass through the interferometric filter crystal (IFC), but in different spatial modes, where dedicated laser beams prepare the required spectral features (transparency window or AFC). Spectral filters at $600\,\mathrm{ nm}$ (width $20\,\mathrm{nm}$) are placed on both arms before the photons are fiber coupled to the single photon detectors (silicon SPD).
(b) Hyperfine splitting of the first sub-levels (0) of the ground $^{3}$H$_{4}$ and the excited $^{1}$D$_{2}$ manifolds of Pr$^{3+}$ in \YSO.
(c) Temporal pulse sequence for the AFC-DLCZ protocol.
$T_S $ ($T_{AS}$) is the time separation between a Stokes (anti-Stokes) photon detection and the write (read) pulse. The insets show the effect of the AFCs in the IFC on the Stokes (orange) and anti-Stokes (red) photons.
(d) Sketch of the Stokes-anti-Stokes coincidence histogram vs $T_S + T_{AS}$ when the IFC is prepared with an AFC in each photon arm, with equal transmission and echo probability. }
\label{setup}
\end{figure*}

In this paper, we use the AFC-DLCZ protocol to create entanglement in time between a single photon and a single collective spin excitation in a REID memory crystal (MC), in the photon counting regime.
The matter state is transfered on demand onto a single photon, and the photonic qubits are analyzed in Franson like interferometers implemented with another REID crystal. The entanglement is demonstrated by observing high-visibility two-photon interference in different bases and by violating a Bell inequality.

In the AFC-DLCZ protocol, a write pulse resonant with an AFC structure with comb spacing $\Delta$ is used. Excited atoms will then start to dephase due to the inhomogeneous broadening. Spontaneously emitted Stokes photons are collected between the write pulse and the corresponding AFC echo, which appears at time $\tau_{MC}=1/\Delta$. The Stokes photons emitted at different times are correlated to independent spin waves leading to entanglement in time between the Stokes photons and the stored spin waves. The joint light matter state can be written to first order as:
\begin{equation}
\mid \Psi_{S,SW} \rangle=\int{(\mathds{1} + \rho(T_S)a_S^\dagger (T_S)a_{SW}^\dagger (T_S))\mid 0_{S}, 0_{SW} \rangle}dT_S
\end{equation}
where $a_S^\dagger$ ($a_{SW}^\dagger$) is the creation operator for the Stokes photon (spin wave), $T_S$ is the time between the write pulse and the Stokes emission and $ \rho(T_S)$ the temporal dependency of the wavefunction \cite{Sekatski2011}.

A strong resonant read pulse can be sent at a later time to excite the spin wave back to the excited states, which, after a finite rephasing time, leads to collective emission of an anti-Stokes photon. Due to the fixed rephasing time of the excited state, the anti-Stokes emission time is correlated with the Stokes emission time following $T_S + T_{AS}$ = $\tau_{MC}$, where $T_{AS}$ is the time between the read pulse and the anti-Stokes photon emission.
In the ideal case of unity read-out efficiency, the joint state of the Stokes anti-Stokes photons  $\mid \Psi_{S,AS} \rangle$ can then  be written as:
\begin{equation}
\int{( \mathds{1} + \rho(T_S)a_S^\dagger (T_S)a_{AS}^\dagger (\tau_{MC}-T_S))\mid 0_{S}, 0_{AS} \rangle}dT_S
\end{equation}
where ($a_{AS}^\dagger$) is the creation operator for the anti-Stokes photon. Our device therefore acts as a source of entangled photons with embedded quantum memory.

\begin{figure*}
	\centering{\includegraphics[width=2\columnwidth]{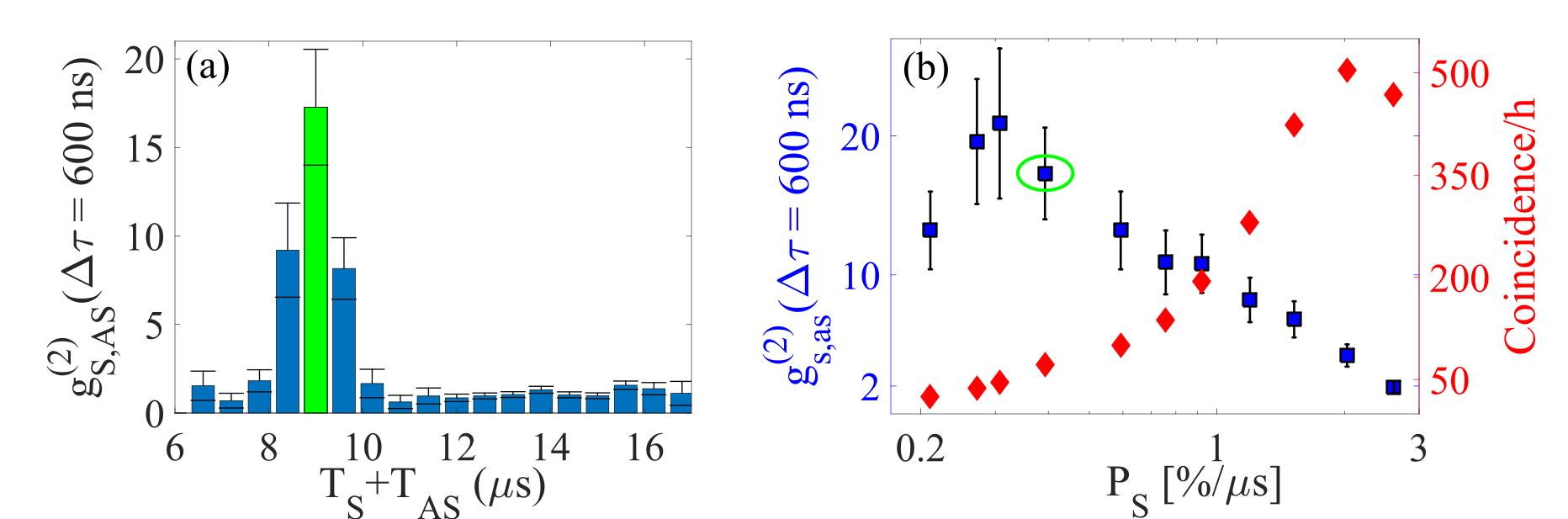}}
	\caption{(a) Typical $g^{(2)}_{S,AS}$ histogram as a function of the $T_S+T_{AS}$ time, with a time-bin size of $600\,\mathrm{ns}$. The write pulse power $P_w$ is $90\,\mu\mathrm{W}$, corresponding to a Stokes creation probability $P_S = 0.4\,\%/\mathrm{\mu s}$. (b) $g^{(2)}_{S,AS}$ (blue squares) and coincidence rate (red diamonds) vs Stokes creation probability. Both quantities refer to a time-bin size of $600\,\mathrm{ns}$. The errorbars are calculated assuming Poissonian statistics. }
\label{coincidence}
\end{figure*}

In this experiment, we use two \PrYSO\, (Pr:YSO) crystals, a memory crystal (MC) and an interferometric filter crystal (IFC), both cooled down to $3.5\,\mathrm{K}$ \cite{Kutluer2017} in a closed loop cryostat. This material offers long coherence times \cite{Heinze2013}, high storage and retrieval efficiencies \cite{Hedges2010}, and prospect for on-chip integration \cite{Corrielli2016,Seri2018}. A sketch of the relevant experimental setup and the energy level scheme of Pr$^{3+}$ in YSO are shown in panels (a) and (b) of Fig. \ref{setup}. We tailor the $1/2 _{g} - 3/2 _{e}$ transition of the MC as an AFC structure with $\tau_{MC} = 9 \,\mathrm{\mu s}$, while the $3/2 _{g} $ state is emptied to store the single spin excitation (see \cite{Seri2017} for more details). We prepare the AFC structures every cryostat cycle ($1\,\mathrm{Hz}$ rate).

We then start to send Gaussian write pulses resonant to the AFC at a rate of 3.7 KHz (1100 pulses per AFC preparation).  We detect the Stokes photons in a $4\,\mathrm{\mu s}$ window starting $1\,\mathrm{\mu s}$ after the write pulse (temporal sequence in Fig. \ref{setup}(c)). As discussed in \cite{Kutluer2017}, the number of modes stored is given by the ratio between the Stokes photon detection window and the duration of the Stokes photon itself. The latter is set by the duration of the write pulse (700 ns of full-width-half-maximum) resulting in 5 distinguished temporal Stokes modes. The Stokes detection mode is set to an angle of about 3 degrees in the backward direction with respect to the write mode, to minimize the leakage noise from the write pulse. Conditional on a Stokes photon detection, we send the Gaussian read pulse, counter-propagating to the write mode and delayed by $16\,\mathrm{\mu s}$.
As a consequence of the phase matching conditions, the Stokes and anti-Stokes photons are emitted in opposite directions. The anti-Stokes detection gate is finally opened for about $10\,\mathrm{\mu s}$. The average storage time in the spin state is $\overline{\tau_S} = 13\,\mathrm{\mu s}$.
Directly after the emission, the Stokes photons are steered to the IFC, where a $2\,\mathrm{MHz}$-wide transparency window is prepared, to filter the photons emitted through the decay to hyperfine ground levels other than the $3/2_g$. The anti-Stokes photons are temporally gated with an acousto-optic modulator (AOM) before traveling through the IFC where another transparency window is created in a different spatial mode to suppress the coherent and incoherent noise deriving from the read pulse \cite{Gundogan2015}.
AFCs can also be created in the IFC, which will serve as qubit analyzers for the photons (see below).  Both Stokes and anti-Stokes photons are detected with single photon counters (SPD) and their arrival time is saved to reconstruct coincidence histograms.

\begin{figure*}

\centering{\includegraphics[width=2\columnwidth]{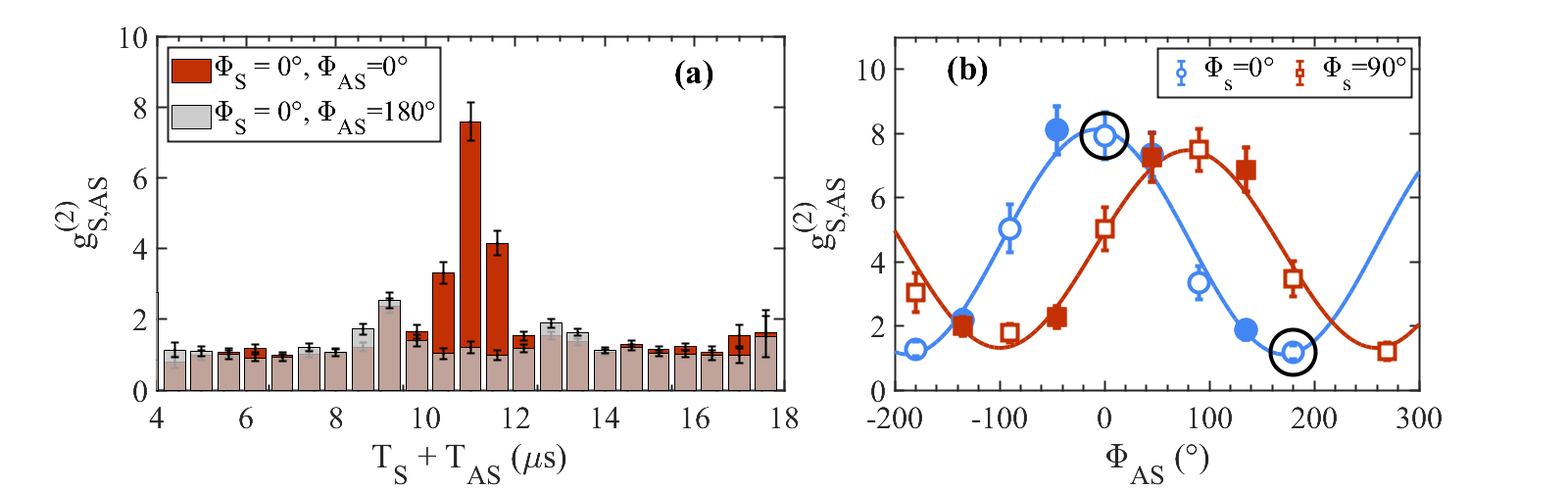}}

\caption{(a) Examples of  $g^{(2)}_{S,AS}$ between Stokes and anti-Stokes photons when an AFC with $\tau _{IFC} = 2\,\mathrm{\mu s}$ is prepared in both the Stokes and the anti-Stokes mode in the IFC. The cases of constructive (darker bars) and destructive (lighter bars) interference are shown. For both measurements, the integration time is $10$ hours, approximately. (b) Interference fringes measured by tuning the frequency of the anti-Stokes filter AFC (i.e. by tuning $\Phi _{AS}$) in two different bases, selected by changing the frequency of the filter AFC for the Stokes photon. The circled points are the one related to Fig.(3). The filled points are those used to calculate the S parameter (integration time 6.5 hours per point). The remaining data point are the result of 5.5 hours of integration each.}
\label{interf:g2}
\end{figure*}
We first verify that the Stokes and anti-Stokes photons are emitted in pairs, and assess the cross-correlation function $g^{(2)}_{S,AS} = \frac{p_{S,AS}}{p_S \cdot p_{AS}}$, where $p_{S,AS}$ is the probability to detect a coincidence between a Stokes and an anti-Stokes photon and $p_S$ ($p_{AS}$) is the probability to detect single Stokes (anti-Stokes) photon.

Fig. \ref{coincidence}(a) shows the measured $g^{(2)}_{S,AS}$ histogram fixing the time-bin size to $\Delta t = 600\,\mathrm{ns}$. This measurement is taken with a write pulse power of $P_W = 90\,\mathrm{\mu W}$, corresponding to a total Stokes creation probability $P_S = 1.6\,\%$ ($P_S=0.4\,\%/\mu s$). We observe a clear peak at $T_S + T_{AS} = 9\,\mathrm{\mu s}$, that represents the correlated Stokes-anti-Stokes pairs, featuring a maximum of $g^{(2)}_{S,AS}=17.3 \pm 3.3$. This is widely above the classical limit of $2$ fixed by the Cauchy-Schwarz inequality, assuming thermal statistics for the Stokes and anti-Stokes fields, as predicted for the DLCZ protocol in the ideal case \cite{Kuzmich2003}. The efficiency to retrieve an anti-Stokes photon, conditioned on a Stokes detection, is about $1.6\,\%$ mostly limited by the read pulse transfer efficiency, uncorrelated background detection in the Stokes mode, and the rephasing efficiency of the AFC and spin-wave decoherence (see \cite{Kutluer2017} and Appendix).

Fig. \ref{coincidence}(b) shows the coincidence count rate and $g^{(2)}_{S,AS}(600\,\mathrm{ns})$ as a function of the Stokes probability $P_S$ (see Appendix for a characterization of $P_S$ vs $P_W$).
As expected, the coincidence rate increases with the Stokes probability. Nonetheless this increases the probability of multiple spin-excitations and creates a noise component proportional to $P_S$, therefore reducing the $g^{(2)}_{S,AS}$ \cite{Laplane2017}. At the lowest $P_S$ values, the cross-correlation decreases because the Stokes photon rate becomes comparable to the noise.The correlation between Stokes and anti-Stokes photons remains largely above the classical limit until $P_S = 2\,\%/\mathrm{\mu s}$, corresponding to $P_W$ of almost $1\,\mathrm{mW}$. However, in the following experiments, we maintain $P_W = 90\,\mathrm{\mu W}$ to guarantee a good balance between coincidence rate and cross-correlation value. Note that for similar value of $g^{(2)}_{S,AS}$, we achieve a coincidence count rate 8 times higher than in our previous demonstration \cite{Kutluer2017}.

To demonstrate energy-time entanglement, measurements in complementary time bases are required. This can be achieved by sending optical fields in unbalanced interferometers serving as time-bin analyzers, as suggested by Franson \cite{Franson1989}. In our case, we use the IFC as a time-bin analyzer by preparing an AFC with storage time of $\tau_{IFC} = 2\,\mathrm{\mu s} $ in both spatial modes. The AFC structure acts on the single photons as a beam splitter with a delay line in one output, i.e. a part of an unbalanced Mach-Zehnder interferometer \cite{Clausen2011, Jobez2015a, Maring2017}. This provides a convenient and robust time-bin analyzer \cite{Jobez2015a}, without the need of phase stabilizing interferometers with several hundred meters path length difference. In the IFC, each Stokes and anti-Stokes photon can be either transmitted (Early time bin E) or stored in the AFC and retrieved as an AFC echo after a time $\tau_{IFC}$ (Late time bin L). The phase $\Phi_{S}$ ($\Phi_{AS}$) between the early and late time-bin can be tuned by changing the center frequency of the AFC with respect to the Stokes (anti-Stokes) photons \cite{Afzelius2009}. A phase shift of $2\pi$ is achieved with a frequency detuning of $\Delta$.

The coincidence histogram between Stokes and anti-Stokes photon detections after the time-bin analyzers, as a function of the $T_S + T_{AS}$ time will be thus composed of three peaks. One corresponds to the coincidences between transmitted Stokes and anti-Stokes  (labeled $\mid E_S E_{AS} \rangle$ in panel (d) of Fig. \ref{setup}) and it thus lays at $T_S + T_{AS} = \tau_{MC}$. One builds up with the coincidences between Stokes and anti-Stokes photons when both undergo AFC storage in the IFC, $\mid L_S L_{AS}\rangle$. Consequently, it will appear at $T_S + T_{AS} = \tau_{MC} + 2\cdot \tau_{IFC}$. The central peak featured in Fig. \ref{setup}(d) at $T_S + T_{AS} = \tau_{MC} + \tau_{IFC}$ is the sum of two contributions: the coincidences between transmitted Stokes photons and stored anti-Stokes photons ($\mid E_S L_{AS} \rangle$) and those between stored Stokes photons and transmitted anti-Stokes photons ($\mid L_S E_{AS} \rangle$). If these two processes are indistinguishable and coherent (which requires e.g. equal AFC echo efficiency  in the IFC for both photons), they will be able to interfere. By selecting only the central peak, the correlation between Stokes and anti-Stokes photons can be interpreted as coming from the postselected entangled state
$\frac{1}{\sqrt{2}} (\mid E_S L_{AS} \rangle + e^{i\theta}\mid L_S E_{AS} \rangle )$, where $\theta = \Phi _{S} - \Phi _{AS}$ \cite{Clausen2011}. This photonic state results from the postselected  light-matter entangled state
$\frac{1}{\sqrt{2}} (a_S^\dagger (T) a_{SW}^\dagger (T)+a_S^\dagger (T') a_{SW}^\dagger (T'))\mid0_S0_{SW}\rangle$
where $T$ and $T'$
are separated in time by $\tau_{IFC}$, set by the analyzing interferometer. In our experiments, we tailor the finesse of the AFC structures in the IFC such that the amplitude of the transmitted and stored pulses are comparable (approximately $30\,\%$ of the input pulses).

Fig. \ref{interf:g2}(a) shows examples of $g^{(2)}_{S,AS}$ between Stokes and anti-Stokes photons when both pass through an AFC in the IFC. The constructive (dark bars) and destructive (light bars) interference cases are reported as obtained by fixing the Stokes phase shift $\Phi _{S}=0^{\circ}$, and the anti- Stokes phase shift $\Phi _{AS}$ to $0\,^{\circ}$ and  $180\,^{\circ}$, respectively.  As expected, the two histograms differ in the area around $T_S + T_{AS} = 11\,\mathrm{\mu s}$.
The value of  $g^{(2)}_{S,AS}$ for constructive interference is 7.6$\pm$0.5, lower than the value measured before the time-bin analyzers. This is due to the fact that there is an intrinsic loss in the IFC ($\eta_{IFC}=30\,\%$), and that the noise from different temporal modes is summed up.

In Fig. \ref{interf:g2}b, we show the results of two photon interference measurements in different bases, obtained by tuning the anti-Stokes phase shift for two different values of $\Phi _{S}$. The visibility is $(75.9\pm4.6)\,\%$ for $\Phi _{S} = 0\,^{\circ}$ and $(70.1\pm4.4)\,\%$ for $\Phi _{S} = 90\,^{\circ}$ showing evidence of entanglement.
To further assess quantum entanglement, we perform an experiment probing the violation of the Clauser, Horne, Shimony, and Holt (CHSH) inequality \cite{Clauser1969a}. We measure thus the coincidence histograms in 16 different settings  to calculate the $S$ parameter as
\begin{equation}
S = E(\alpha, \beta) + E(\alpha', \beta) + E(\alpha, \beta') - E(\alpha', \beta'),
\label{CHSH}
\end{equation}
where $\alpha$ and $\alpha'$ ($\beta$ and $\beta'$) are two different phase choices for the Stokes (anti-Stokes) photons arm.
The different terms of the eq. \ref{CHSH} are build from the coincidences C in a time bin $\Delta t = 600\,\mathrm{ns}$ around $T_S+T_{AS} = 11\,\mathrm{\mu s}$ as follows:
\begin{eqnarray*}
E(\alpha, \beta) =\\ \frac{C(\alpha, \beta)+C(\alpha + \pi, \beta+ \pi)-C(\alpha, \beta+ \pi)-C(\alpha + \pi, \beta)}{C(\alpha, \beta)+C(\alpha + \pi, \beta+ \pi)+C(\alpha, \beta+ \pi)+C(\alpha + \pi, \beta)}
\label{terms}
\end{eqnarray*}
We fix the phase values to $\alpha = 0\,^{\circ}$, $\alpha' = 90\,^{\circ}$, $\beta = 45\,^{\circ}$, and $\beta' = 135\,^{\circ}$ and obtain $S = 2.15 \pm 0.07$ that surpasses the classical bound of 2 by more than 2 standard deviations.

These measurements show that the Stokes and anti-Stokes photons are entangled in time. Consequently, as the conversion spin wave to anti-Stokes photon is a local operation, this also demonstrates entanglement between a photon and the spin-wave stored in the crystal.

We have shown that a solid-state device can emit pairs of entangled photons with an embedded quantum memory for one of the photons. Moreover, the quality of the entanglement is high enough to enable a violation of a Bell inequality, making our device suitable for applications in quantum communication. With improved performances (see Appendix), this device could be an important resource for the implementation of temporally multiplexed quantum repeaters. It could also serve as a platform for investigating high-dimensional entanglement in time between light and matter.

\textbf{Acknowledgments.} We acknowledge financial support by the European Union via the Quantum Flagship project QIA (820445), by the Gordon and Betty Moore foundation through Grant GBMF7446 to HdR, by the Spanish Ministry of Economy and Competitiveness (MINECO) and Fondo Europeo de Desarrollo Regional (FEDER) (FIS2015-69535-R), by MINECO Severo Ochoa through grant SEV-2015-0522, by Fundaci\'o Cellex, and by CERCA Programme/Generalitat de Catalunya. HdR aknowledges useful discussions with Nicolas Sangouard.
\section{Appendix}
\subsection{A. Experimental Setup}

\begin{figure}
	\centering{\includegraphics[width=1.3\columnwidth]{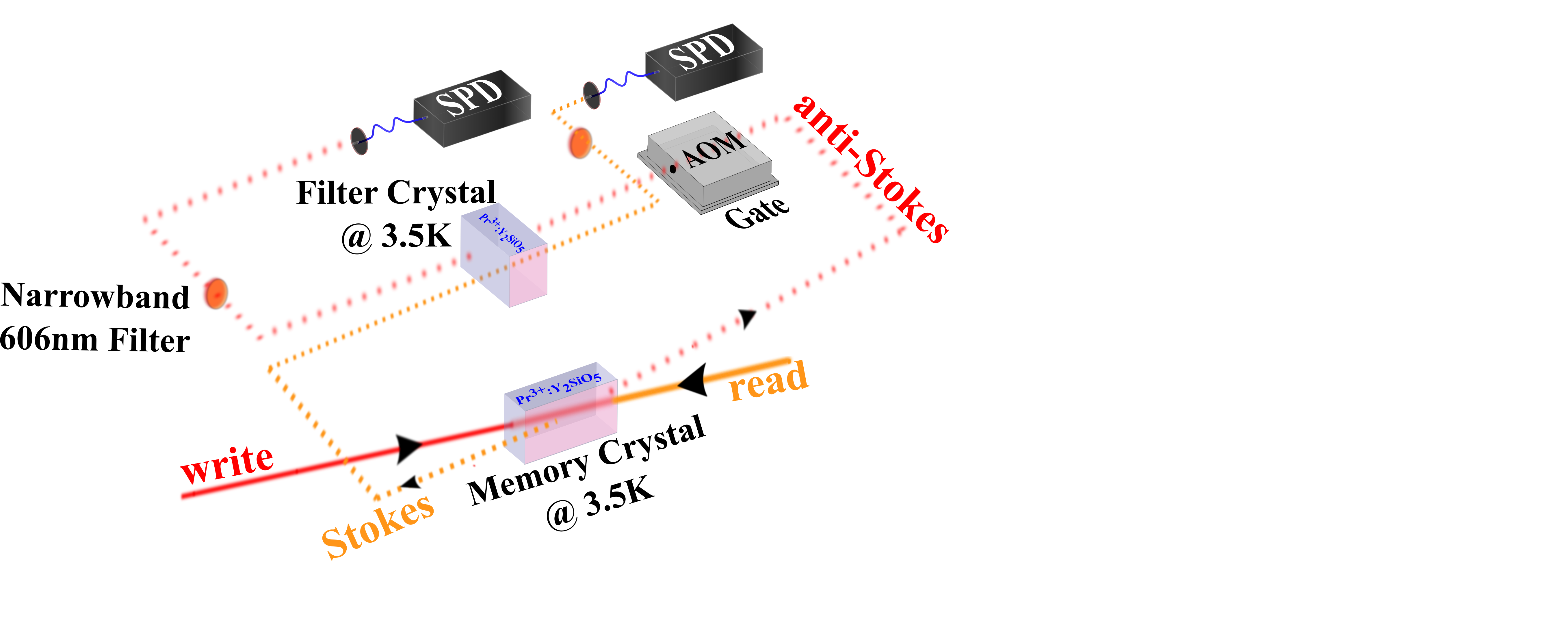}}
	\caption{Sketch of the experimental setup. } 
\label{setup}
\end{figure}

The experimental setup is depicted in detail in Fig. \ref{setup}. The \PrYSO\, crystals used as quantum memory (MC) and interferometric filter (IFC) are hosted in a cryostat (Montana Cryostation) and cooled down at a temperature of $3.5\,\mathrm{K}$.  The AFC in the memory crystal is prepared with the read pulse mode. 
After the AFC preparation, we wait $145\,\mathrm{ms}$ in order to skip the noisy part of the cryostat cycle. 
The write and read pulses applied in the DLCZ protocol are polarized parallel to the D$_{2}$ crystal axis to maximize the interaction.
The write pulses are $700\,\mathrm{ns}$ long (FWHM).
The read pulses have Gaussian amplitude modulation and a maximum power of $30\,\mathrm{ mW}$. They are $1\,\mathrm{ \mu s}$ long and frequency chirped of $800\,\mathrm{kHz}$ with a hyperbolic tangent waveform.
Both Stokes and anti-Stokes photons pass through the interferometric filter crystal (IFC), but in different spacial modes (about $4\,\mathrm{mm}$ apart), where dedicated laser beams prepare the required spectral features (transparency window or AFC). 
Before the IFC, the anti-Stokes photons are temporally gated by an acousto-optic modulator (AOM). Narrow-band filters at $600\,\mathrm{ nm}$ (width $20\,\mathrm{nm}$) are placed in both arms before the photons are fiber coupled to the single photon detectors (silicon SPD, efficiency $\eta_D = 50\,\%$). The total transmission in the Stokes (anti-Stokes) arm, from the cryostat to the detector, is typically $59\,\%$ ($56\,\%$), which takes into account the passive losses of the optical elements and the residual absorption in the IFC.

\subsection{B. Stokes creation probability}
\label{PS}
\begin{figure}
	\centering{\includegraphics[width=0.9\columnwidth]{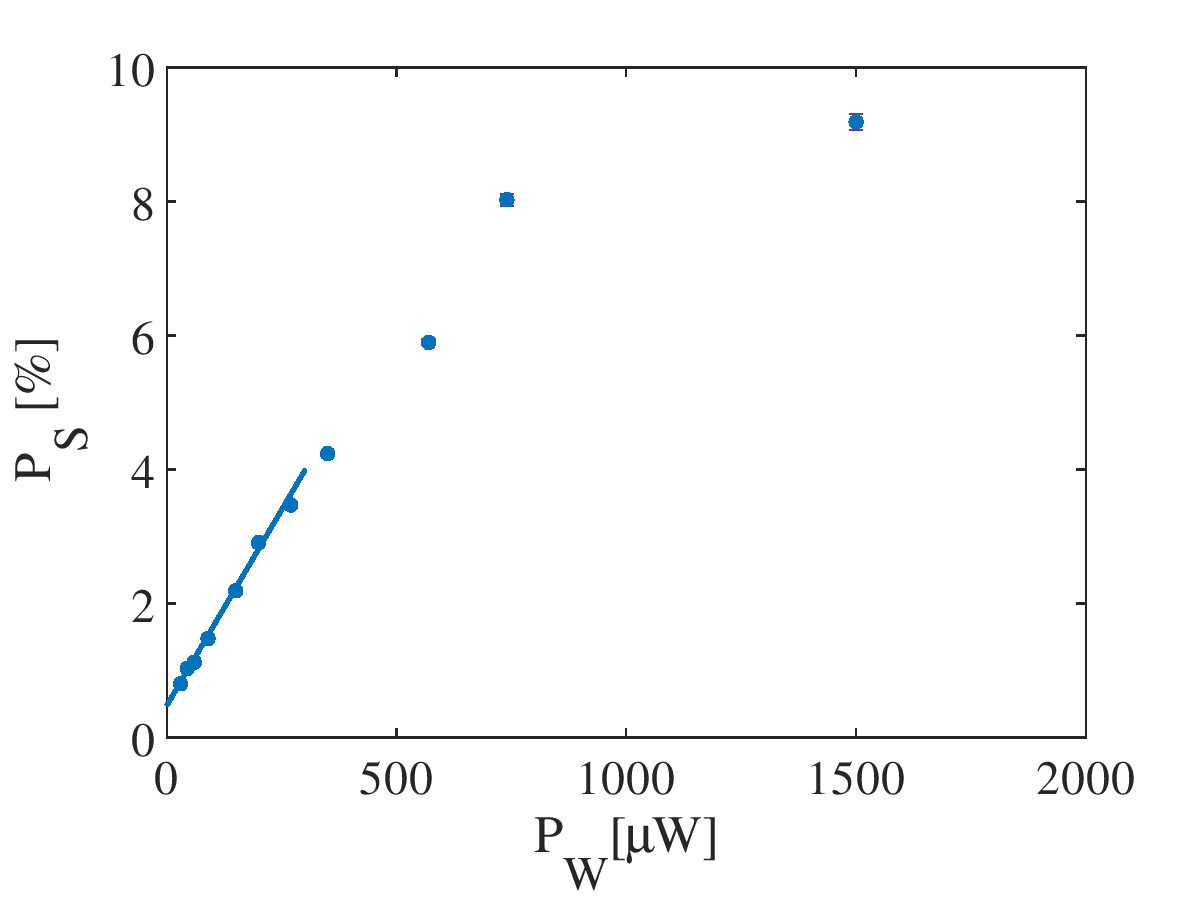}}
	\caption{Stokes creation probability, $P_S$, as a function of the write pulse power, $P_W$. A linear fit of the data points at lower $P_W$ is also shown. The fitting curve intercepts the y axis at $P_S = 0.5\,\%$. } 
\label{Ps_vs_Pw}
\end{figure}

A characterization of the Stokes creation probability, $P_S$, as a function of the write pulse power is shown in Fig. \ref{Ps_vs_Pw}. The scaling is linear, with the exception of the highest write pulse power investigated where, we believe, the sequence of write pulses might destroy the comb structure. Notably, the linear interpolation of the data points at lower write pulse powers highlights a residual background at $P_W = 0\,\mathrm{\mu W}$. This corresponds to a $P_S$ of about $0.5\%$, which is a remarkable portion of the total Stokes probability, especially at lower write pulse power. We believe that it might be due to stray light reaching the Stokes detector. The fact that a considerable portion of the Stokes detections does not in fact correspond to a spin wave in the crystal largely affects our anti-Stokes readout efficiency.

\subsection{C. Coincidence histograms}

Each detection of a Stokes photon is followed by 10 unconditional write-read pairs, with $3.8\,\mathrm{kHz}$ repetition rate, in order to detect the accidental counts for estimating the second-order cross correlation function between Stokes and anti-Stokes photon. 
An example of Stokes-anti-Stokes coincidence histogram as a function of the $T_S + T_{AS}$ time is shown in Fig. \ref{coincidence}. The red bars plot the Stokes-anti-Stokes coincidence counts in the conditional storage trials and the solid curve is the average of the coincidence counts in the following $10$ unconditional trials. The latter represents the accidental counts. 
We observe a clear peak at $T_S + T_{AS} = 9\,\mathrm{\mu s}$, that represents the correlated Stokes-anti-Stokes pairs. Besides this main peak, the histogram also includes a broader peak at longer times. 
This is a leakage of the third AFC echo of the write pulse in the anti-Stokes mode, likely due to scattering in the memory crystal, that cannot be filtered by the IFC because it is resonant with the transparency window (see energy level scheme in Fig. 1 of the main text). 
However, as it is present in both curves, it is washed out in the calculation of the $g^{(2)}_{S,AS}$. 

\begin{figure}
	\centering{\includegraphics[width=0.8\columnwidth]{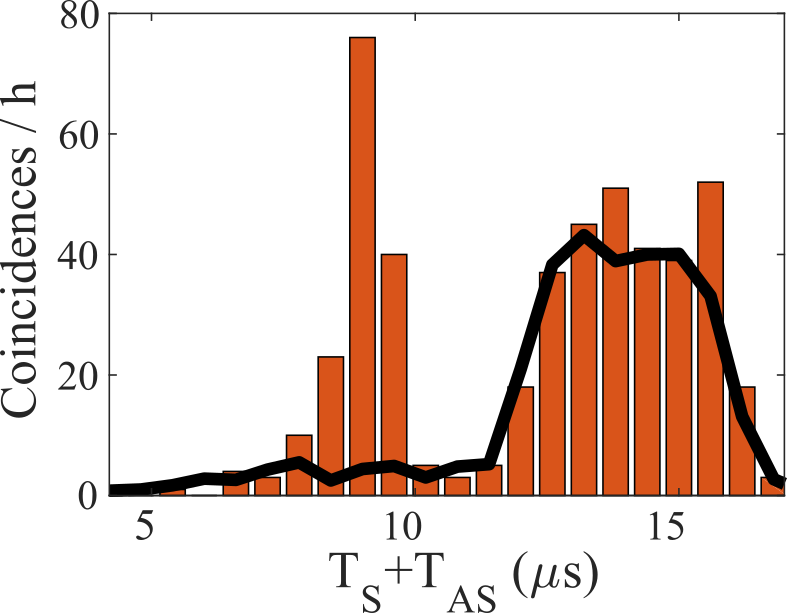}}
	\caption{Time resolved Stokes-anti-Stokes coincidence histogram (red bars) plotted together with the accidental coincidence histogram (solid curve) for $P_W = 90\,\mathrm{\mu W}$, corresponding to $P_S = 1.6\,\%$. } 
\label{coincidence}
\end{figure}

\begin{figure}
	\centering{\includegraphics[width=0.8\columnwidth]{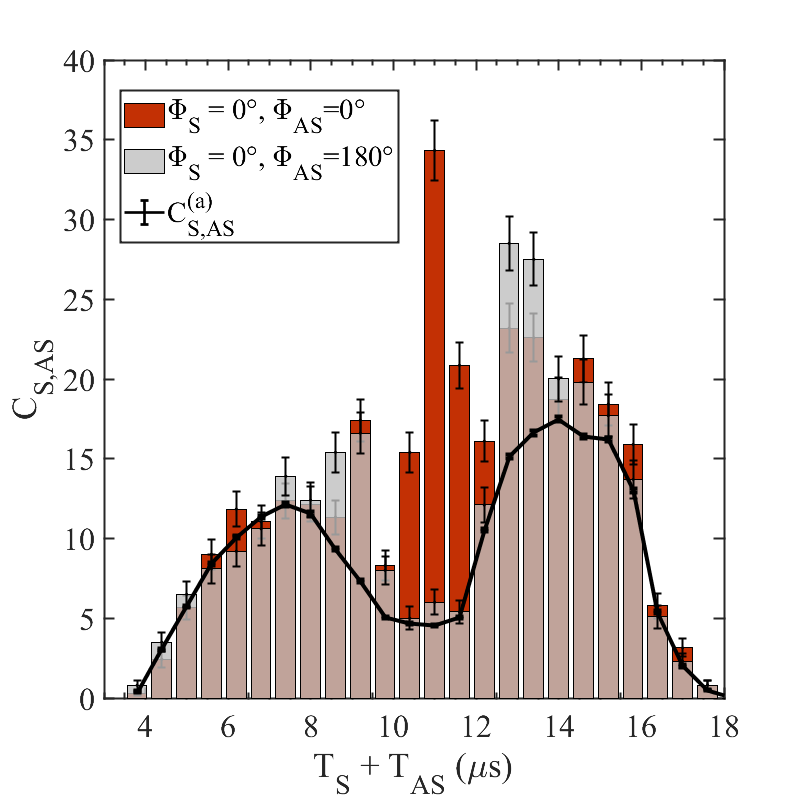}}
	\caption{Time resolved coincidence histogram (red bars) plotted together with the accidental coincidence histogram (solid curve). The number of coincidences per hour is plotted. The write pulse power is again $P_W = 90\,\mathrm{\mu W}$ ($P_S = 1.6\,\%$). } 
\label{int_coincidence}
\end{figure}

Fig. \ref{int_coincidence} reports examples of coincidence histograms between Stokes and anti-Stokes photons $C_{S,AS}$ when an AFC with $\tau _{IFC} = 2\,\mathrm{\mu s}$ is prepared in both the Stokes and the anti-Stokes mode in the IFC. The cases of constructive (darker bars) and destructive (lighter bars) interference are shown, along with the accidental coincidence counts $C^{(a)}_{S,AS}$ for the constructive case (solid curve). The constructive and destructive interference measurements mainly differ in the region around $11\,\mathrm{\mu s}$, that is $\tau_{MC} + \tau_{IFC}$. For both measurements, the integration time is $10$ hours, approximately.
In the accidental coincidence histogram two broad peaks appear. We attribute them to AFC echoes originated in the IFC of the leakage of the write pulse and its multiples echoes in the anti-Stokes mode. However, as in the case above, when calculating the $g^{(2)}_{S,AS}$ these peaks disappear as they are present in all the coincidence histograms.

\subsection{D. Analysis of the CHSH inequality violation}

\begin{figure}
	\centering{\includegraphics[width=0.8\columnwidth]{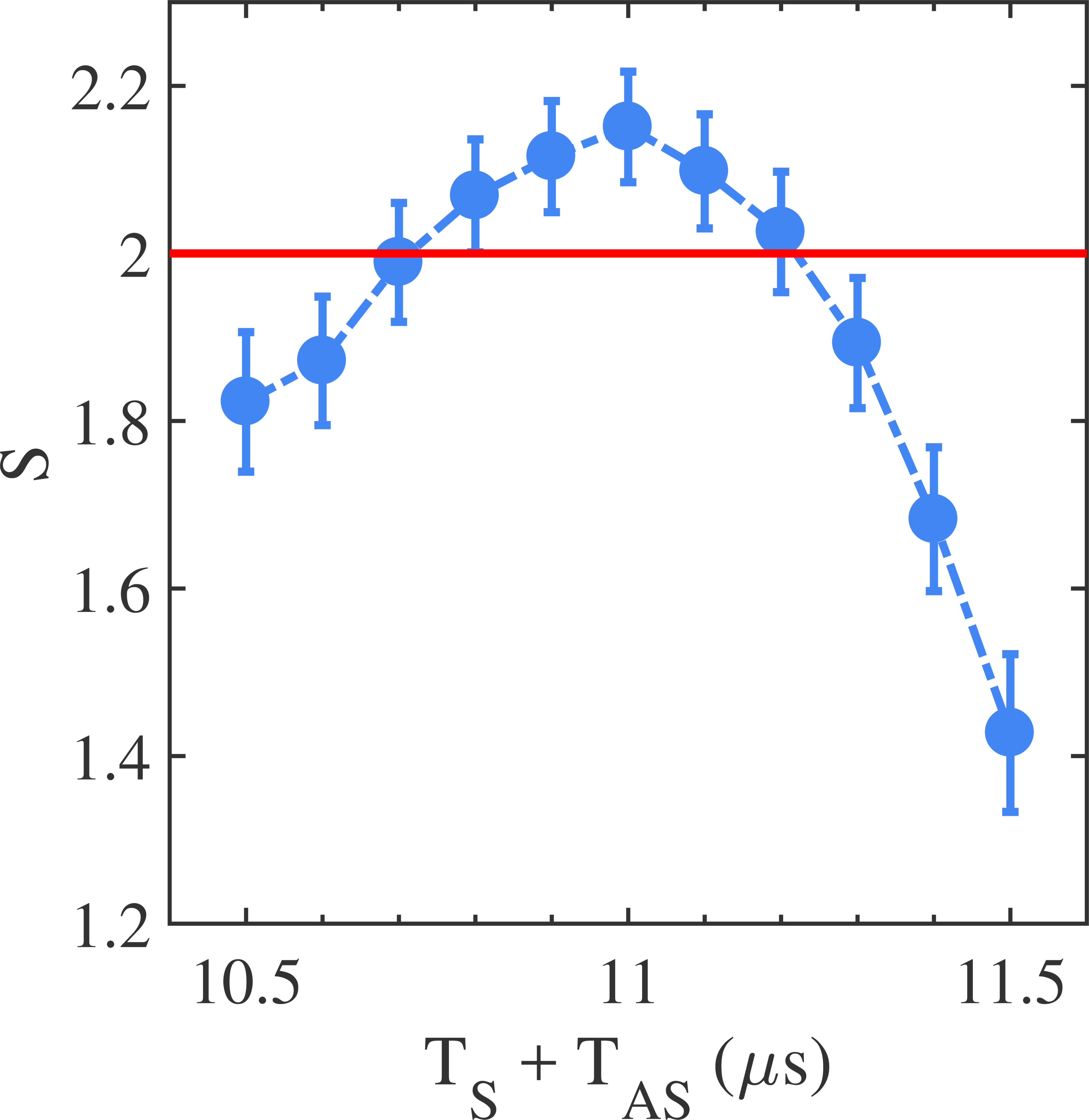}}
	\caption{S parameter as a function of the coincidence window position, when this is moved around the peak at $\tau_{MC} + \tau_{IFC}$. } 
\label{position}
\end{figure}

To obtain the maximum violation of the CHSH inequality, we do a characterization of the $S$ parameter (as defined in the main text). Figure \ref{position} shows the behavior of $S$ when the $600\,\mathrm{ns}$-coincidence window is moved around the peak at $T_S + T_{AS} = 11\,\mathrm{\mu s}$ (see as an example the coincidence histograms of Fig. \ref{int_coincidence}). We observe a maximum violation exactly at $T_S + T_{AS} = 11\,\mathrm{\mu s}$, that corresponds to $\tau_{MC} + \tau_{IFC}$. 
In this case, the $E(\alpha,\beta)$ for $\alpha=0^\circ$, $\alpha'=90^\circ$, $\beta=45^\circ$ and $\beta'=135^\circ$ are 
\begin{gather*}
E(0^\circ,45^\circ)=0.50 \pm 0.03\\ 
E(90^\circ,45^\circ)=0.56 \pm 0.03\\
E(0^\circ,135)^\circ)=-0.59 \pm 0.03\\
E(90^\circ, 135^\circ)=0.50 \pm 0.04.
\end{gather*} 

From these values, we obtain $S = 2.15 \pm 0.07$. If compared to the maximum value of $2 \cdot \sqrt{2}$, we estimate a visibility of $V_{CHSH} = 76.5 \pm 2.5$, compatible, within error bars, with the average value of the visibilities measured from the interference fringes of Fig. 3(b) of the main text. 

\begin{figure}
	\centering{\includegraphics[width=0.8\columnwidth]{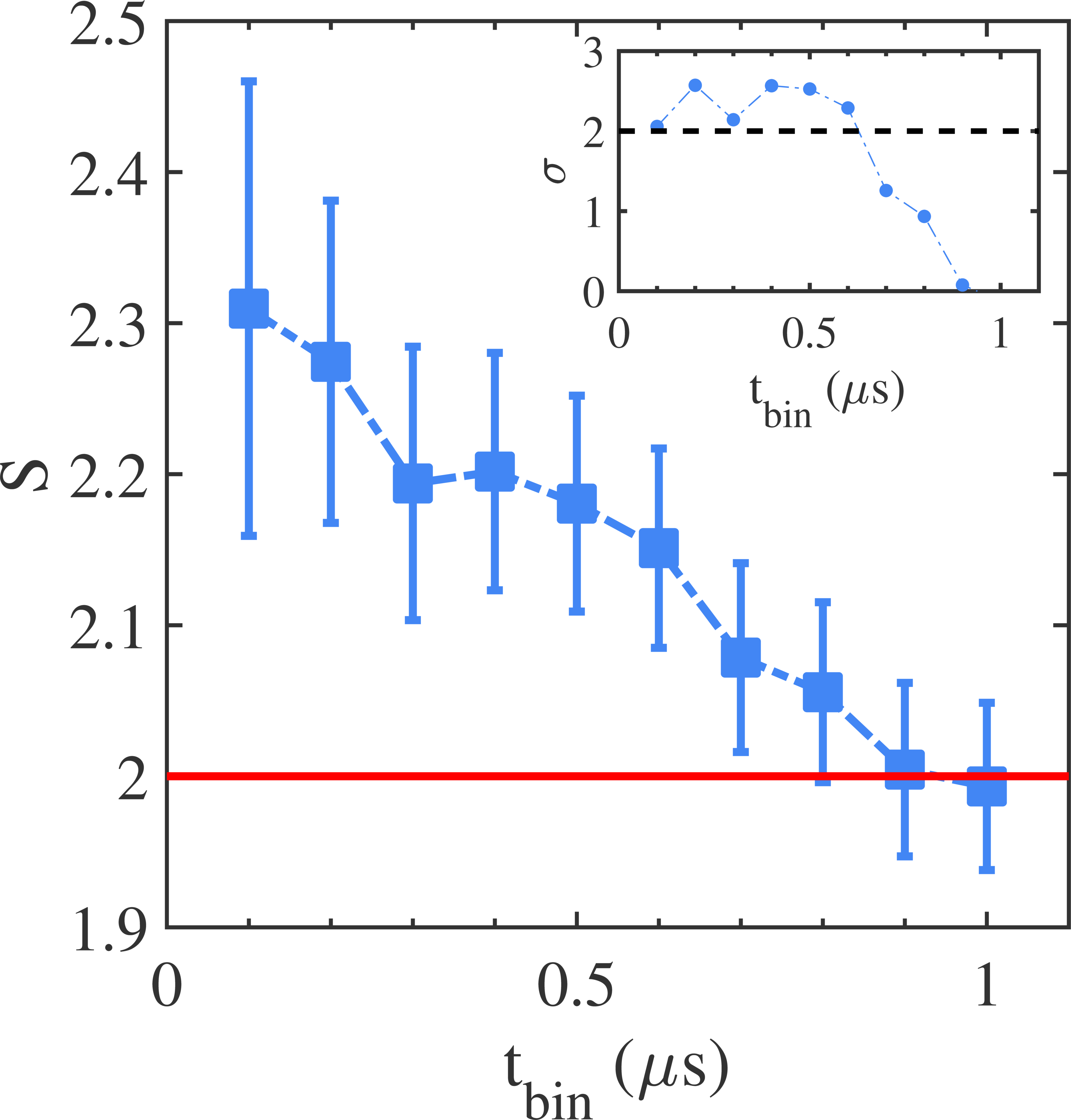}}
	\caption{S parameter as a function of the time-bin size. The coincidence window is fixed at $\tau_{MC} + \tau_{IFC}$ Inset: violation of the CHSH in terms of standard deviations vs time-bin size. } 
\label{timebin}
\end{figure}

However, there is a finite range of positions in which the CHSH inequality is clearly violated. 
We then fix the window position at the maximum and analyze the $S$ parameter as a function of the time-bin size, as depicted in Fig. \ref{timebin}. The inset shows the corresponding violation of the CHSH inequality in terms of standard deviations. 

\subsection{E. Discussion on storage time and readout efficiency}

In this experiment we achieve a $g^{(2)}_{S,AS} = 17.3 \pm 3.3 $ and an anti-Stokes retrieval efficiency (conditioned on a Stokes detection) $\eta_{R} = 1.6\,\%$ when $P_W = 90\,\mathrm{\mu W}$, corresponding to a Stokes creation probability of $P_S = 1.6\,\%$. 
We note that these values are comparable to those reported in our previous realization of the AFC-DLCZ protocol \cite{Kutluer2017}. However, in the present work the average storage time in the spin state $\overline{\tau_S}$ is more than a factor $2$ longer and the coincidence count rate has been increased by a factor 8. 

With the current storage time in our memory (currently limited by the spin-inhomogeneous broadening), we could demonstrate light-matter entanglement over distances of a few kilometers. For the transmission of the photon in optical fibers, quantum frequency conversion to telecom wavelengths would be required, as was recently shown in ref. \cite{Maring2018} for this wavelength. Longer storage times can be readily achieved by implementing spin-echo sequences to overcome the inhomogeneous broadening (leading to $T_2$ =500 $\mu s$ for our crystal). Ultimately, the application of a suitable magnetic field and of dynamical decoupling techniques \cite{Laplane2017} may prolong the storage time up to tens of seconds in our crystal \cite{Heinze2013} and up to a few hours in Europium doped samples \cite{Zhong2015a}.  
 
\begin{figure}
	\centering{\includegraphics[width=1.0\columnwidth]{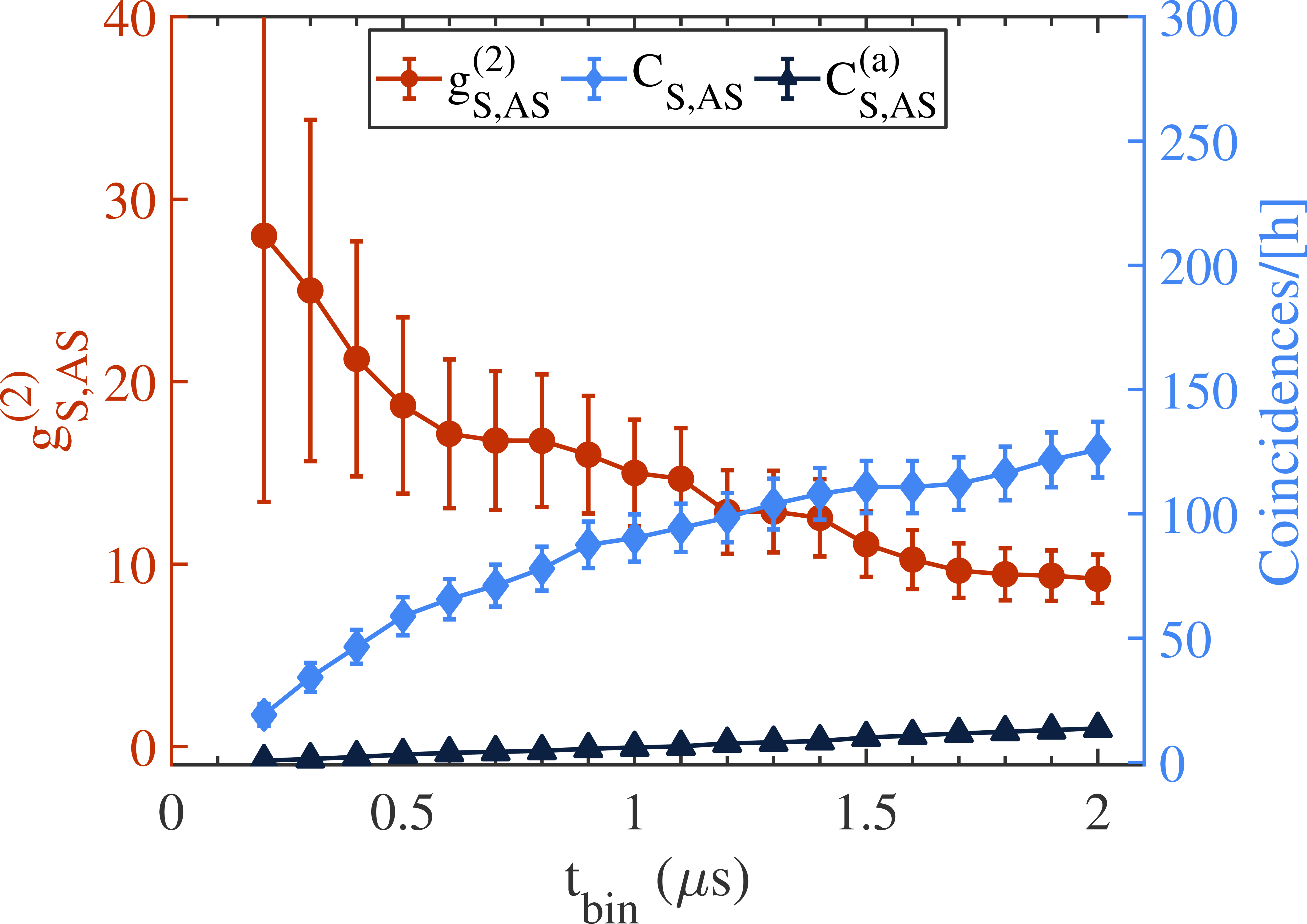}}
	\caption{The cross-correlation function $g^{(2)}_{S,AS}$ (red dots), the coincidences per hour $C_{S,AS}$ (blue diamonds) and the accidental coincidences per hour $C^{(a)}_{S,AS}$ (black triangles) as a function of the coincidence detection window width $t_{bin}$. } 
\label{g2_coincidences}
\end{figure}

The current coincidence count rate is mostly limited by the low retrieval efficiency and signal-to-noise ratio and by the repetition rate of the experiment. On the one hand, higher signal-to-noise ratio would allow increasing $P_S$, and thus the coincidence rate, while keeping the same value for the the $g^{(2)}_{S,AS}$. This can be achieved by increasing the read out efficiency and suppressing the background noise. As discussed in \cite{Kutluer2017}, the efficiency could be increased by improving the AFC preparation, by a better control of the spin decoherence as well as by using a longer crystal with higher optical depth and/or by embedding the crystal in a low finesse cavity \cite{Afzelius2010a}. Furthermore, background noise suppression would also allow us to widen the coincidence window, leading to a higher coincidence rate, as show in Fig. \ref{g2_coincidences}. On the other hand, the repetition rate $R$ for experiments aiming at entangling remote crystals in a heralded fashion is limited by the communication time between the two memories, $R=L/c$ (e.g. 4 kHz for L=50 km). The coincidence count rate could however be enhanced by increasing the number of modes. The number of temporal modes can be increased by using longer storage time in the excited state in the memory crystal (several tens of $\mu s$ are possible in our crystal). Additional multiplexing can be achieved by implementing frequency and spatial mode multiplexing, e.g. using an integrated approach \cite{Seri2018}.

\bibliography{qpsa}  
\bibliographystyle{prsty}

\end{document}